# Tracer Kinetic Models as Temporal Constraints during DCE-MRI reconstruction

Authors: Sajan Goud Lingala[1], Yi Guo[1], R. Marc Lebel[2], Yinghua Zhu[1], Yannick Bliesener[1], Meng Law[3], Krishna S. Nayak[1,3]

[1] Ming Hsieh Department of Electrical Engineering, University of Southern California, Los Angeles, California

[2] GE Healthcare Applied Sciences Laboratory, Calgary, Canada

[3] Department of Radiology, University of Southern California, Los Angeles, California

Corresponding Author:

Sajan Goud Lingala, PhD

Post-doctoral Research Associate

Ming Hsieh Department of Electrical Engineering

University of Southern California

lingala@usc.edu

**(ABSTRACT – 200-word limit)**

**Purpose:** To apply tracer kinetic models as temporal constraints during reconstruction of under-sampled dynamic contrast enhanced (DCE) MRI.

**Methods:** A library of concentration v.s time profiles is simulated for a range of physiological kinetic parameters. The library is reduced to a dictionary of temporal bases, where each profile is approximated by a sparse linear combination of the bases. Image reconstruction is formulated as estimation of concentration profiles and sparse model coefficients with a fixed sparsity level. Simulations are performed to evaluate modeling error, and error statistics in kinetic parameter estimation in presence of noise. Retrospective under-sampling experiments are performed on a brain tumor DCE digital reference object (DRO) at different signal to noise levels (SNR=20-4ID0) at (k-t) space under-sampling factor (R=20), and 12 brain tumor in-vivo 3T datasets at (R=20-40). The approach is compared against an existing compressed sensing based temporal finite-difference (tFD) reconstruction approach.

**Results:** Simulations demonstrate that sparsity levels of 2 and 3 model the library profiles from the Patlak and extended Tofts-Kety (ETK) models, respectively. Noise sensitivity analysis showed equivalent kinetic parameter estimation error statistics from noisy concentration profiles, and model approximated profiles. DRO based experiments showed good fidelity in recovery of kinetic maps from 20-fold under-sampled data at SNRs between 10-30. In-vivo experiments demonstrated reduced bias and uncertainty in kinetic mapping with the proposed approach compared to tFD at R>=20.

**Conclusions:** Tracer kinetic models can be applied as temporal constraints during DCE-MRI reconstruction, enabling more accurate reconstruction from under-sampled data. The approach is flexible, can use several kinetic models, and does not require tuning of regularization parameters.

**INTRODUCTION**

Dynamic Contrast Enhanced MRI (DCE-MRI) is a powerful technique that provides a quantitative measure of vessel permeability and interstitial volumes. In the brain, it characterizes the blood brain barrier (BBB) leakiness, which has proven to be valuable in several applications [1]. These include assessing conditions with large BBB breakdown such as gradation of brain tumors [2,3], multiple sclerosis lesions [4,5], and conditions with subtle and chronic BBB breakdown such as diabetes [6], and Alzheimer's disease [7]. Outside the brain, DCE-MRI has applications in cancer assessment and therapeutic monitoring in several body parts including breast [8,9], prostate [10], and liver [11].

DCE-MRI involves a challenging trade-off between the achievable spatial resolution, temporal resolution, and volume coverage. Acceleration strategies that exploit redundancies along the time dimension have shown significant potential to improve these trade-offs. These include early schemes such as view-sharing [12,13,14], highly constrained back projection (HYPR) [15], and more recently compressed sensing [16,17,18,19,20,21]. Several sparsifying spatio-temporal transforms have been proposed including spatio-temporal wavelet transform, spatio-temporal finite-difference, temporal Fourier transform. A major challenge with these "off-the-shelf" object models is that the modeling assumptions do not fit the data, which limits the achievable acceleration rates. Data-driven schemes that learn sparse representations from the data have been proposed [22-25], and have shown to out perform off-the shelf transforms. However, these are often associated with highly non-convex optimization. Furthermore, image reconstruction with existing transforms involves tuning one or more regularization parameters, which poses challenges to the standardization of these methods.

In this manuscript we explore the use of physical tracer kinetic models for constrained reconstruction. This approach has been used extensively in dynamic positron emission tomography (PET) imaging [26,27,28,29], and has recently been adapted in MRI for the applications of relaxometry [30,31,32,33], perfusion [34,35], and diffusion imaging [36,37]. Broadly, these methods can be classified into methods based on direct reconstruction of parameters from under-sampled data, or methods that use representations derived from parametric models as constraints in image reconstruction.

We propose a model-constrained approach for DCE-MRI, where established contrast-agent kinetic models are used as temporal constraints. From a specific kinetic model, and a physiological range of kinetic parameters, we construct a library of concentration vs. time profiles. Kinetic model specific temporal basis functions are derived from the library using the k-singular value decomposition (k-SVD) algorithm [38]. Through noise-less and noise-based simulations, we deduce a relation between the sparsity parameter in k-SVD and the complexity level of the kinetic model. We design a constrained reconstruction method where the kinetic model-based temporal bases are used to constrain the recovery of concentration v.s time profiles from under-sampled (k-t) data. We utilize an iterative multi-scale optimization algorithm for improved robustness to undesirable local minima solutions.

The proposed approach is unique because the constraints are designed based on contrast-agent TK models that are routinely used during post-processing. The main difference between this approach and direct parametric reconstruction is that the reconstruction of time profiles is decoupled from parameter estimation. This allows for flexibility in dealing with complex non-linear kinetic models, and also is compatible with AIF parametric model forms. Furthermore, since the sparsity parameter, is fixed *a priori*, the proposed approach does not require any tuning of free parameters (e.g. regularization parameters). The flexibility allows for its potential utility in DCE-MRI of most organs and disease conditions. In this work, we demonstrate effectiveness with both the Patlak and extended Tofts-Kety (ETK) models, and demonstrate application to brain tumor assessment.

# METHODS

## Tracer kinetic model-based temporal bases:

A library of concentration vs. time profiles $Z_{lxN}$ is simulated using a kinetic model, an arterial input function (AIF), and a broad physiologic range of kinetic parameters (Fig.1). $l$ denotes the number of profiles in the library; and $N$ denotes the number of time instances. For the ETK model [39], we used the range: $K^{trans} = 0 - 0.8$ min$^{-1}$ in steps of 0.01 min$^{-1}$, $v_p = 0 - 60$ % in steps of 1 %, $v_e = 0 - 100$ % in steps of 1 % to yield a library of size $lxN = 494100 \times 50$. Similarly, for the Patlak model [40], we used the range: $K^{trans} = 0 - 0.8$ min$^{-1}$ in steps of 0.01 min$^{-1}$, $v_p = 0 - 60$ % in steps of 1 % to yield a library size $lxN = 4941 \times 50$. A population based AIF was used [41]. The settings of the Parker model that specifies the population based AIF were the same as described in [41]. The range of kinetic parameters was motivated by brain tumor DCE literature [1], which suggests 0-0.34 min$^{-1}$ for $K^{trans}$, 0-60% for $v_p$ assuming hematocrit of 0.4, and 0-100% for $v_e$. We expanded the $K^{trans}$ range by ~2.5x, and used the full range for $v_p$ and $v_e$ to ensure conservative coverage of the kinetic parameter space. The k-SVD dictionary-learning algorithm [38] is then used to reduce the large library to a smaller dictionary of temporal basis functions (denoted by $V_{rxN}$). k-SVD represents any time profile in Z, for instance the $p^{th}$ row of Z, $z_p(t)$, as a sparse linear combination of basis functions $v_i(t)$ from V:

$$\underbrace{z_p(t)}_{1 \text{X} N} \approx \underbrace{u_p V_{r \text{X} N}}_{z_p^{q-sp}(t)} \quad s.t., \quad \left\| \underbrace{u_p}_{r \text{x} 1} \right\|_0 \leq q; \qquad [1]$$

where $r$ denotes the number of basis functions in V, and is chosen as $r = 100 \ll l$. $q$ is the sparsity parameter. $||u_p||_0$ denotes the $l_0$ norm of the vector $u_p = \{u_1, u_2, \dots, u_r\}$. $z_p^{q-sp}(t)$ denotes the $q$-sparse projection of $z_p(t)$ onto V. k-SVD jointly estimates the sparse coefficient matrix $U_{lxr}$ and the dictionary $V_{rxN}$ as:

$$\{\hat{U}_{lxr}, \hat{V}_{rxN}\} = \min_{U,V} \sum_{p=1}^{l} \|z_p(t) - u_p V\|_2^2; \text{ s.t.}, \|u_p\|_0 \leq q; \qquad [2]$$

where $u_p$ denotes the $p^{th}$ row of U.

**<u>Image Reconstruction:</u>**

We pose the estimation of the concentration vs. time profiles $C_{MxN}$ ($M$ - number of pixels; $N$ - number of time frames), and the sparse coefficient matrix $U_{Mxr}$ from under-sampled k-t space data (**b**) as:

$$\min_{C,U} \underbrace{\|A(C) - b\|_2^2}_{\text{data consistency}} ; s.t., \underbrace{C = UV; \|u_p\|_0 \leq q; p = \{1,2, \ldots, M\}}_{\text{TK model constraint}}; \qquad [4]$$

C contains the concentration v.s time profile $c(\mathbf{x}, t)$ for every pixel $\mathbf{x} \in (x, y)$ stacked row wise. $\gamma(\mathbf{x}, t)$ are constrained to be a q-sparse linear combination of the kinetic model-derived temporal bases in $V_{rxN}$. The operator $A = F_u\left(S_m(T(C))\right)$ denotes the forward model which maps C to the measured multi-coil (k,t) data. $F_u$ denotes the Fourier Transform operator on a specified (k-t) under-sampling pattern. $S_m$ contains the coil sensitivity maps. In this work, the sensitivity maps are estimated from time averaged data using the standard root-sum-of-squares method and are assumed to capture object phase. T is an operator that relates the concentration profile to the signal intensity profile $s(\mathbf{x}, t)$ by the steady state spoiled gradient echo (SPGR) equation:

$$s(\mathbf{x}, t) = T(c(\mathbf{x}, t)) = \frac{M_0(\mathbf{x}) \sin\alpha \left(1 - e^{-TR[R_1(\mathbf{x},0) + c(\mathbf{x},t)r_1]}\right)}{1 - \cos\alpha (e^{-TR[R_1(\mathbf{x},0) + c(\mathbf{x},t)r_1]})}$$

$$+ \left[s(\mathbf{x}, 0) - \frac{M_0(\mathbf{x}) \sin\alpha \left(1 - e^{-TR[R_1(\mathbf{x},0)]}\right)}{1 - \cos\alpha (e^{-TR[R_1(\mathbf{x},0)]})}\right]; \quad [5]$$

where $r_1$ is the contrast agent relaxivity, $TR$ is the repetition time, $\alpha$ is the flip angle, $R_1(\mathbf{x}, 0)$ and $M_0(\mathbf{x})$ are respectively the pre-contrast $R_1$ (reciprocal of $T_1$) and the equilibrium longitudinal magnetization. $s(\mathbf{x}, 0)$ is the pre-contrast first frame, which is fully sampled. The bracketed term in the second row of [5] resolves differences between the pre-contrast signal $s(\mathbf{x},0)$ and the predicted pre-contrast signal based on the baseline $R_1(\mathbf{x},0)$ and $M_0(\mathbf{x})$ maps (from a separate $T_1$ mapping acquisition). Similarly, the operation of mapping concentration profile from the signal intensity profile can be expressed as [42]:

$$c(\mathbf{x},t) = T^{-1}(s(\mathbf{x},t))$$

$$= \frac{-\frac{1}{TR}\ln\left[\frac{1-\left(\frac{s(\mathbf{x},t)-s(\mathbf{x},0)}{s(\mathbf{x},0)\sin\alpha}+\frac{1-e^{-TR[R_1(\mathbf{x},0)]}}{1-\cos\alpha(e^{-TR[R_1(\mathbf{x},0)]})}\right)}{1-\cos\alpha\left(\frac{s(\mathbf{x},t)-s(\mathbf{x},0)}{s(\mathbf{x},0)\sin\alpha}+\frac{1-e^{-TR[R_1(\mathbf{x},0)]}}{1-\cos\alpha(e^{-TR[R_1(\mathbf{x},0)]})}\right)}\right] - R_1(\mathbf{x},0)}{\Re 1} \quad;[6]$$

We solve [4] by alternately (a) updating U using orthogonal matching pursuit (OMP) sparse projection [38,43], and (b) updating $C$ by enforcing consistency with acquired data. To be robust to spurious local minima, we use an iterative multi-scale minimization approach, where we solve the problem at a coarser spatial resolution during the initial iterations and as the iterations proceed, we gradually update the resolution to its full resolution. This is achieved by multiplication of spatial Fourier Transform of $s(\mathbf{x},t)$ by a 2D Gaussian filter $(G(k_\sigma))$ specified by filter width $k_\sigma$; where $k_\sigma$ is initialized to 0.1 percent of $k_{max}$, and gradually updated to 100 percent of $k_{max}$, where $k_{max}$ specifies the extent of k-space coverage. This heuristic strategy is used in several non-convex problems such as in image registration [44], and more recently in MR-fingerprinting [45,46]. Starting with an initial guess obtained from $\Gamma_{init} = A^H(\mathbf{b})$, we iterate until a stopping criterion of $\left\|\frac{|C_i - C_{i-10}|}{|C_i|}\right\|_2^2 < \varepsilon = 0.01$ or until the maximum number of iterations of 150 are achieved. After reconstructing $\hat{C}$, we estimate the kinetic parameters by fitting the estimated concentration profiles to the kinetic model using Rocketship [47]. The pseudo code of the algorithm is shown below.

______________________________________________________

**Initialization:** $C_{init} = A^H(\mathbf{b})$; $\tau_{init} = 0.001 * k_{max}$

*while* $k_\sigma < k_{max}$

- For all time frames, spatially blur $s(\mathbf{x}, t)$ by the 2D Gaussian filter $G(k_\sigma)$
- $k_\sigma = k_\sigma * 2$
- Map signal intensity profiles to concentration profiles: $c(\mathbf{x}, t) = T^{-1}(s(\mathbf{x}, t))$;

while $\left( \left\| \frac{|C_i - C_{i-10}|}{|C_i|} \right\|_2^2 < \varepsilon \right)$

- *TK Model constraint update*
  - OMP update of $u_p$, s.t, $u_p V \approx c(\mathbf{x}, t)$; $\|u_p\|_0 \leq q; p \in \{1,2,\ldots,M\}$;
- Map the k-sparse projected concentration profiles to the signal intensity profiles: $s(\mathbf{x}, t) = T(u_p V)$;
- *Data consistency update*
  - Compute $\hat{s}(\mathbf{k}, t_j) = F_u \left( S_m (s(\mathbf{x}, t_j)) \right)$; for $j \in \{1,2,\ldots,N\}$; and insert the measured data at the sampling locations $\hat{s}(\mathbf{k}_u, t_j) = \mathbf{b}$;
- Map the above k-t space data to the concentration time profiles: $c(\mathbf{x}, t) = A^T \left( \hat{s}(\mathbf{k}, t_j) \right)$;

*end*

end

---

## Simulations

The sparsity parameter $q$ in [2] is determined based on simulation studies with the Patlak and e-Tofts models. Noise-less simulations are performed and the mean approximation error $\mu_{err} = \frac{1}{l} \sum_{p=1}^{l} \left\| z_p(t) - z_p^{q-sp}(t) \right\|_2^2$, and maximum approximation error: $max_{err} = max_{p=1}^{l} \left\| z_p(t) - z_p^{q-sp}(t) \right\|_2^2$ are computed for different values of $q$.

Noise based simulations were performed for broad ranges of kinetic parameter values to a) determine any systematic bias and uncertainty in the kinetic parameter space that may be induced by sparsity based modeling of the concentration time profiles, and b) to deduce the correspondence between the sparsity level (q) and the kinetic model.

Noisy concentration profiles were obtained as:

$$z_p^n(t) = z_p(t) + \mathrm{n}(t); p = \{1, \ldots, l\}; [3]$$

where n(t) denotes i.i.d. white Gaussian noise with zero mean and 0.005 standard deviation, which was chosen to match the typical signal-to-noise ratio (SNR) from *in vivo* brain DCE-MRI data acquired at our institution on a 3T commercial system with an eight-channel head array coil. Noise in the concentration time profiles was assumed to be additive i.i.d. Gaussian. Concentration time profiles are real valued (negative values can occur in the presence of noise), and have a one-to-one mapping with real-valued signal intensity time profiles in the forward model. At the flip angles commonly used in DCE-MRI, this mapping is approximately linear. Monte-Carlo simulations with 500 realizations of $\mathrm{n}(t)$ were performed to evaluate the bias and uncertainty in estimating kinetic parameters from a) the noisy profiles $z_p^n(t)$, and b) the $q$ – sparse projections of $z_p^n(t)$ on V: $z_p^{n,q-sp}(t)$.

We performed covariant error analysis for two parameters ($K^{trans}$, $v_p$) with the Patlak and the ETK model over a broad range of kinetic parameters. With both the models, we evaluated the bias and uncertainty in estimating $K^{trans}$ and $v_p$ before and after q-sparse projections. With the ETK model, for simplicity, we focus only on analysis in a two dimensional space with a fixed ve =0.6. The open-source Rocketship package [47] was used for TK parameter estimation.

**Evaluation with a digital reference object**

An anatomically-realistic brain tumor DCE-MRI digital reference object (DRO) was generated based on the method and data described in [48]. Briefly, the population based AIF with the Parker model, known TK parameters, the extended Tofts model, and the steady state spoiled gradient signal equation was used to generate the dynamic images. We then multiplied by coil sensitivities, took the Fourier Transform, and added realistic complex Gaussian noise to each channel. Coil maps, noise covariance matrix, and the signal to noise (SNR) level were obtained from in-vivo data acquired at 3T. Comparisons were performed at multiple SNR levels of 40, 30, 20; where a SNR = 30 mimicked measurements at 3T.

This phantom data was retrospectively under-sampled using a randomized golden-angle Cartesian (GOCART) sampling pattern, and evaluations in fidelity of the kinetic parameters were performed at under-sampling factor of R=20. GOCART is originally a 3D golden angle Cartesian sampling scheme, with random sampling of the ky-kz phase encode locations along each Cartesian radial spoke. In this study, we perform retrospective under-sampling in the kx-ky plane in a representative slice. This strategy was chosen to simulate $k_y$-$k_z$ under-sampling in prospective acquisitions. The empirical stopping criterion of our algorithm was also evaluated at different SNR levels.

**Evaluation with in-vivo data**

We reviewed 110 fully-sampled DCE-MRI raw datasets from patients with known or suspected brain tumor, receiving a routine brain MRI with contrast on a clinical 3T scanner (HDxt, GE Healthcare, Waukesha, WI). A T1-weighted spoiled gradient echo DCE-MRI sequence was used with scan parameters FOV: 22x22x4.2cm$^3$, spatial resolution: 0.9x1.3x7 mm$^3$, temporal resolution: 5 sec, flip angle=15$^0$, TR/TE=6 ms/1.3 ms. DESPOT1 was performed with flip angles of 2°, 5°, 10°, to compute pre-contrast $T_1$ and $M_0$ maps [49]. The contrast agent, Gadobenate dimeglumine (MultiHance Bracco Inc.) was administered with a dose of 0.05 mMol/kg, followed by a 20 ml saline flush in the left arm by intravenous injection. Of these 110 cases, we identified a cohort of 12 cases, which had different brain tumor characteristics (shape, size, heterogeneity), and also had enhancing tumors of atleast 1 cm (as determined by standard bi-directional assessment) [50]. The demographics of these patients are shown in Table 1, and the post-contrast images (last spatial frame from the DCE-scans) are shown in Figure 2. The protocol was approved by our institutional review board (IRB).

Modeling error in the kinetic parameter space was analyzed based on the fully-sampled reference data. e-Tofts model derived temporal bases were used with a fixed sparsity level of q=3. The e-Tofts model was chosen as it accounts for backflux of contrast from the extravascular space to the plasma, which in turn improves the accuracy of $K^{trans}$ estimation, and has shown to be applicable to brain tumor data [51]. Kinetic parameters estimated from the 3-sparse projected profiles (R=1) were compared against the kinetic parameters estimated from the reference concentration time profiles (R=1).

(k-t) under-sampling was performed retrospectively on fully-sampled raw data using the GOCART (randomized golden angle Cartesian) sampling trajectory [52] at acceleration factors R=20 and R=40. Image reconstruction was performed with the proposed dictionary based approach, and compared with an existing compressed sensing approach that uses a temporal finite difference (tFD) sparsity constraint [19]. e-Tofts derived bases with a fixed sparsity level of q=3 was used in the proposed approach. All the patient datasets were acquired with a fixed injection timing, however timing delays between 5-10 seconds (1-2 frames) existed amongst different patients. As described earlier, a population based AIF with a fixed delay was used to generate the library. Patient specific AIF delays were estimated as described by Lebel et al [53]. Briefly, the k-space origin was frequently sampled, and plotted as a function of time. The region of maximum slope was regressed to the baseline to determine the bolus arrival time. Either padding zeros initially to the acquired data or omitting the last time frames corrected for any delay mismatch to the library.

The regularization parameter for tFD constrained reconstruction was tuned to provide the smallest normalized root mean squared image reconstruction error (nRMSE) in tumor ROIs with respect to the reference fully sampled datasets. All reconstructions were implemented in MATLAB (The MathWorks, Inc., Natick, MA) and executed on an Intel core i7 3.5 GHz machine with 32-GB memory.

The convergence of the proposed multi-scale iterative optimization was evaluated empirically. Reconstruction estimates with different initializations of the concentration profiles were compared: zero filled reconstruction ($\Gamma_{init} = A^H(\mathbf{b})$); low spatial resolution estimate obtained from the center 3x3 window of the k-space data in every time frame ($\Gamma_{init} = \Gamma_{low.res}$); and from the reference fully sampled data ($\Gamma_{init} = \Gamma_{ref}$);

After image reconstruction, the ETK model was used to estimate the kinetic parameters with a population based AIF [41]. Bland-Altman analysis was performed to evaluate systematic bias and uncertainty of the reconstructed $\hat{K}^{trans}$ and $\hat{v}_p$ maps (from the proposed and tFD approaches) with respect to the reference fully sampled $K_{ref}^{trans}$ and $v_{p,ref}$. Comparisons using $v_e$ maps were not considered, as its estimation is associated with high uncertainty with the ETK model [47].

**RESULTS**

**Simulations**

Fig. 3 shows the mean and maximum approximation errors between the concentration vs. time profiles in the library, and the profiles obtained from q-sparse projections onto V at different sparsity levels (q). q-sparse projections of the curves generated from the Patlak and the ETK models are respectively shown in Fig. 2a, and 2b. These curves are chosen to represent different types of tumor enhancement dynamics [54]. With q=1, we observe considerable bias in approximating the TK model generated curves with both the Patlak and the ETK models. However, for the Patlak model, a choice of q≥2 provided excellent agreement with the profiles in the library ($max_{err}/\mu_{err} = 10^{-28}\%/10^{-30}\%$). Similarly, for the e-Tofts model, a choice of q≥3 approximated the profiles in the library with ($max_{err}/\mu_{err} = 2\%/0.008\%$).

Figs. 4 and 5 demonstrates the bias and uncertainty in estimating kinetic parameters in presence of noise. Over a broad range of kinetic parameters, we observe that estimating the kinetic parameters from noisy profiles and the q-sparse projected profiles are equivalent when q≥2 (for the Patlak model), and q ≥ 3 (for the ETK model). Based on these simulations, we fixed q=2 for the Patlak model, and q=3 for the ETK model.

Fig. 6 shows the DRO results evaluating the $K^{trans}$ estimates obtained from reconstructions using the proposed approach (R=20) and reference (R=1) at different SNR levels. According to our empirical stopping criterion, the number of iterations for SNR settings of 40,30,20 respectively was: 122, 128, 134. As expected, we observed faster convergence with higher SNR data. As the SNR is decreased, the $K^{trans}$ estimates depict noise in both the reference and the proposed approach. It can be seen that the proposed approach at R=20 depicts good spatial fidelity of the $K^{trans}$ maps at all the noise levels. This suggests that the reconstruction from the approach was not stuck in a spurious local minima solution at reduced SNR levels.

**Evaluation with in-vivo data**

Fig. 7 shows the evolution of the objective function in [4] as a function of CPU reconstruction time with different initializations of concentration time profiles a) from low-resolution dynamic images $\Gamma_{low.res.}$; b) from zero-filled dynamic images $\Gamma = A^T(b)$; c) from reference dynamic images $\Gamma_{ref}$. The multi-scale optimization gradually updates the complexity of the problem. Due to spatial low-pass filtering, the under-sampling artifacts in initial iterations are considerably reduced making the problem well-posed. $\hat{\Gamma}$ is updated gradually with increasing resolution, as a result of which a monotonic convergence is observed. We empirically found this approach to be robust to local minima; the final solutions were identical with different initializations.

Figure 8 shows retrospective under-sampling comparisons of $K^{trans}$ at R=20. tFD reconstructions resulted in considerable under estimation of $K^{trans}$ in 9 of the 12 cases, while the proposed method was found to be robust to this bias. tFD also relied on adjusting the regularization parameter. In contrast, the proposed parameter free reconstruction provided improved accuracy in $K^{trans}$ mapping. It also provided superior fidelity in maintaining spatial characteristics of the tumors in all cases (eg. depiction of thin tumor boundaries in cases 1 to 5).

Figure 9 shows Bland-Altman plots of the difference between estimated TK parameters (at R=20, R=40) and reference TK parameters on all the 12 cases combined. In comparison to tFD, the proposed approach showed reduced bias and reduced certainty during estimation of $K^{trans}$ and $v_p$. A systematic bias of under estimating $K^{trans}$ and $v_p$ was present in tFD, which is also qualitatively shown in Figure 8.

## DISCUSSION

We have developed a new DCE-MRI reconstruction approach that applies kinetic models as temporal constraints. Based on simulation studies, we deduced a relation between sparsity parameter q in k-SVD to the complexity of the kinetic model. We have demonstrated equivalence of Patlak and ETK models with dictionaries constructed respectively with q=2 and q=3. This approach exploits the smooth time intensity DCE patterns by using temporal basis functions derived from a kinetic model. This is in contrast to generic off-the-shelf transform bases that are blind to the kinetic model behavior of the time intensity profiles. Since the basis functions are designed to mimic only smooth profiles, they are extremely tolerant to noise and under-sampling artifacts, which exhibit rapid temporal oscillations. We also proposed a robust multi- scale iterative optimization algorithm to solve the resulting objective function. We demonstrated empirical robustness to local minima. In-vivo validation with 12 brain tumor cases demonstrated superior recovery performance with the proposed method compared to tFD (reduced bias, uncertainty in kinetic mapping, and better spatial fidelity of kinetic maps) at up to R=40.

The framework can be extended in several ways. A uniform grid of kinetic parameters was used in this study to generate the library of possible concentration profiles from a chosen kinetic model. However, it is possible to perform application-specific discretization of the kinetic parameters to improve sensitivity and accuracy in modeling time curves that lie in a particular zone in the kinetic parameter space.

The framework is flexible to incorporate any kinetic model, which makes it extendable to other body part like breast, liver or prostate, with an appropriate model for specific organ. Complementary constraints such as spatial sparsity could be added to further improve the recovery.

In this feasibility study, we used a population-averaged AIF [41], after accounting for patient-specific AIF delays [53]. Population-averaged AIF's are known to produce a potential bias in the final kinetic maps [55], however in this study, the bias identically affects the reference maps, and maps produced by the proposed reconstruction and the temporal finite difference reconstruction that is used for comparison. The proposed framework can be extended to account for patient-specific AIF variations, e.g. delay and dispersion [56,57], or patient-specific AIF waveforms directly extracted from the data [53].

R2* effects were not included in the forward model [58]. Our DCE-MRI scans were performed with ½ dose (0.05 mmol/kg), and used a short TE of 2ms. We have examined several clinical datasets at our institution and have found phase and R2* effects to be insignificant in tissue and in vessels. We therefore did not consider R2* or off-resonance effects, but these could be easily added to the forward model.

Our clinical DCE-MRI scans had a short scan time of 5 minutes, which was insufficient to recover $v_e$. This is consistent with reports in the literature where estimation of $v_e$ had high uncertainty with short scan times [59]. We have included a range of values for $v_e$ in the library because accounting for backflux is known to improve the estimation of $K^{trans}$ and $v_p$ [39].

In this study, the T1 maps were estimated prior to reconstruction using DESPOT1 with three flip angles [49]. Using fully sampled data, the joint estimation of T1 and kinetic parameter maps has recently been shown to improve accuracy of DCE kinetic parameter maps [60]. An extension of the proposed framework to include joint T1 estimation would warrant the inclusion of multiple T1-based simulated concentration curves in the library, and exploration of superior learning approaches

(alternate to k-SVD) that offer better compression capabilities to efficiently represent a richer library. Furthermore, there could be a number of approaches to improve pre-contrast T1 mapping in a separate step. For instance, increasing the number of flip angle measurements, use of constrained imaging methods (e.g. model-based reconstruction, MR fingerprinting).

The proposed approach has similarities and important distinctions with prior art. Similar to MR-fingerprinting [46,61], our approach exploits physical models for reconstruction. However, it does not modify the acquisition parameter settings. It takes a two-step approach of first reconstructing the concentration time profiles, and then estimating the kinetic parameters in a final step. In comparison to MR-Fingerprinting, our approach is sensitive to motion because the basis functions do not account for motion. However, if reasonable estimates of the motion deformation fields are known or can be estimated from the data, it can be corrected by integration into the forward model [62,63].

Data inconsistencies such as motion, or B1 non-uniformity, may violate the assumption of the appropriateness of the kinetic model on the concentration time profiles. This is equally true with existing compressed sensing methods. However, the framework can seamlessly accommodate prior information in the forward model to improve data consistency (eg. integration of motion maps, B1 maps). The proposed reconstruction assumes the chosen kinetic model to be appropriate to the data. While the kinetic model of choice can be motivated based on the application at hand, the framework is flexible to generate comprehensive libraries from more than one kinetic model. Future application specific studies with chosen kinetic models are however needed to deduce the relation between the complexity of the library and the sparsity parameter (q) during dictionary generation, and subsequently acceleration capabilities.

| Case no. | Age/Sex | Diagnosis |
|---|---|---|
| 1 | 74/M | Glioblastoma |
| 2 | 60/M | Metastatic Melanoma |
| 3 | 44/F | Meningioma |
| 4 | 79/F | Metastatic melanoma |
| 5 | 63/M | Meningioma |
| 6 | 68/M | Glioblastoma |
| 7 | 73/M | Metastatic melanoma |
| 8 | 38/F | Meningioma |
| 9 | 67/M | Renal Cell Carcinoma |
| 10 | 71/M | Pituitary adenoma |
| 11 | 73/F | Meningioma |
| 12 | 54/F | Meningioma |

Table 1: Patient demographic information and diagnosis of the brain tumor cases used in this study.

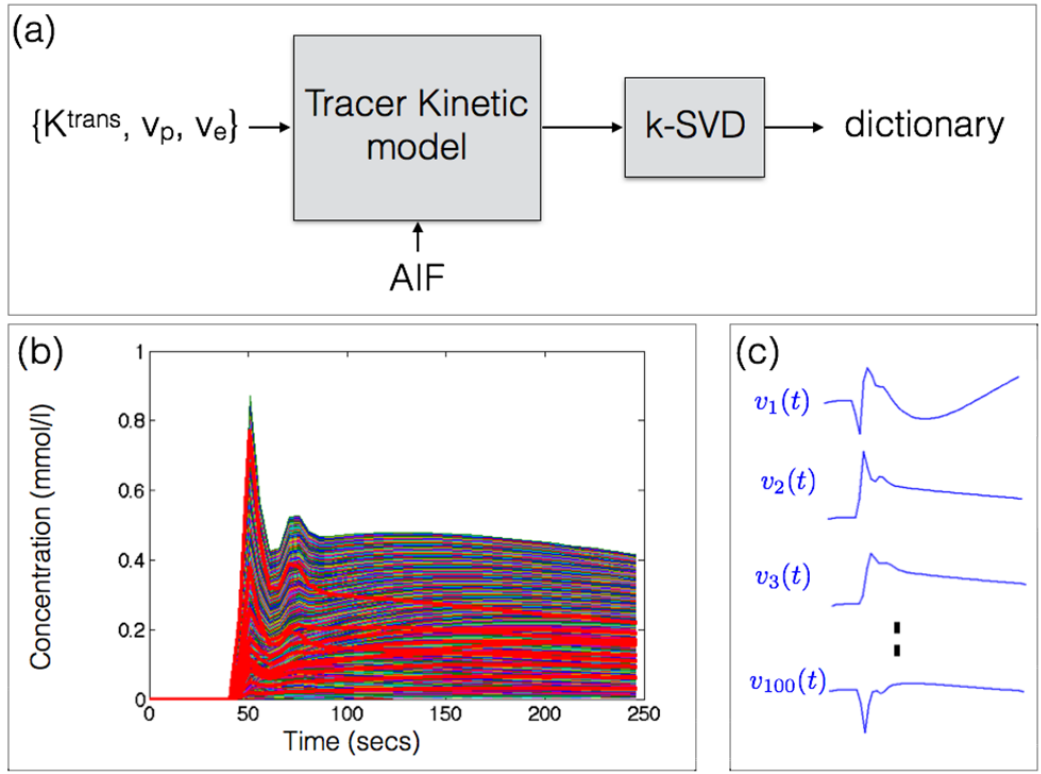

Figure 1: Construction of dictionary of temporal basis functions from a specified tracer kinetic model (a). Based on a physiological range of kinetic parameters and an arterial input function (AIF), a library of concentration v.s time profiles is generated (b). A subset of the profiles in the library are highlighted in red. Using k-SVD, the library is then reduced to a smaller set of temporal basis functions in a dictionary (c). The basis functions generated with the ETK model is shown in (c). The basis functions themselves are not representative of kinetic model profile, and hence can be non-positive. Instead, the linear combination of them is designed to mimic any profile in (b). Approximate MATLAB computational times respectively for generating the library (~400,000 profiles) and learning the dictionary were 11.5 minutes and 3.5 hours.

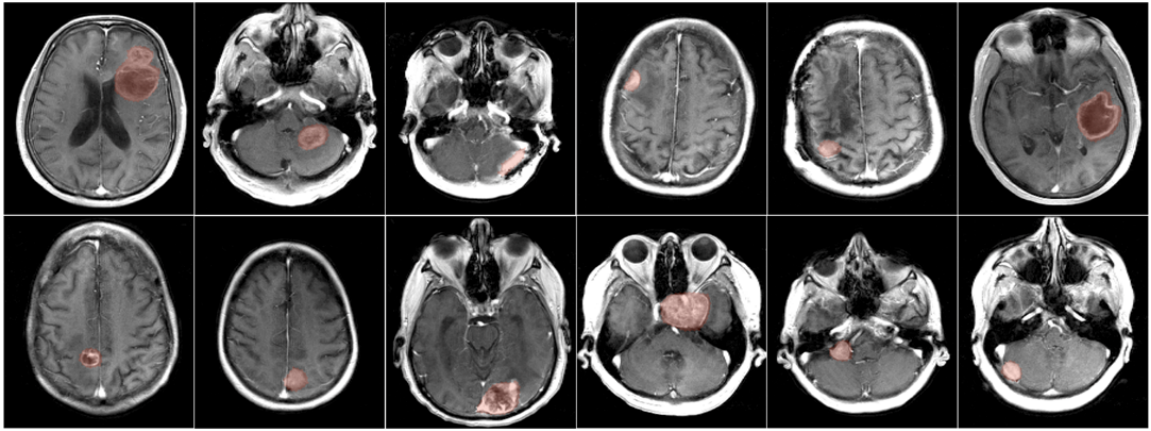

Figure 2: Post-contrast images of 12 brain tumor cases with different brain tumor characteristics (shape, size, heterogeneity). All cases had enhancing tumors of at least 1 cm as determined by standard bi-directional assessment [50]. Fully sampled raw multi-coil (k-t) space data from these patients were used as reference in retrospective under sampling studies. TK parameter estimation was performed in the tumor regions of interests (ROI) as marked by the red shaded regions.

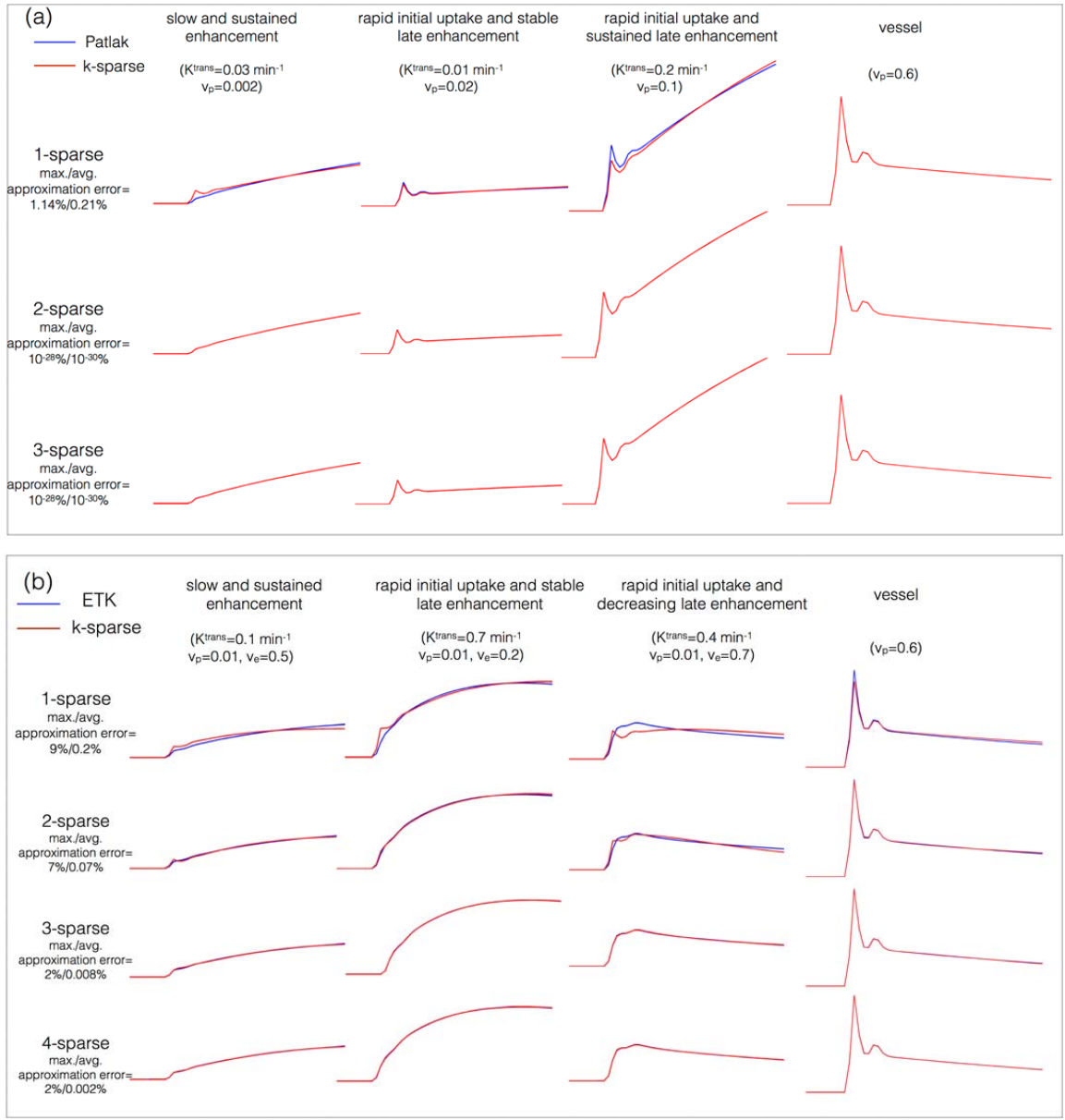

Figure 3: Kinetic model generated concentration v.s time profiles and their representation using k-SVD derived temporal bases. (a) and (b) respectively show representative profiles depicting different tumor enhancement dynamics from the Patlak, and ETK models. The maximum and average approximation errors are evaluated over the physiological range of kinetic parameters. A model-sparsity choice of q=2 was determined to be adequate for the Patlak model ($max_{err}/\mu_{err} = 10^{-28}\%/10^{-30}\%$). Similarly, q=3 was adequate for the ETK model ($max_{err}/\mu_{err} = 2\%/0.008\%$).

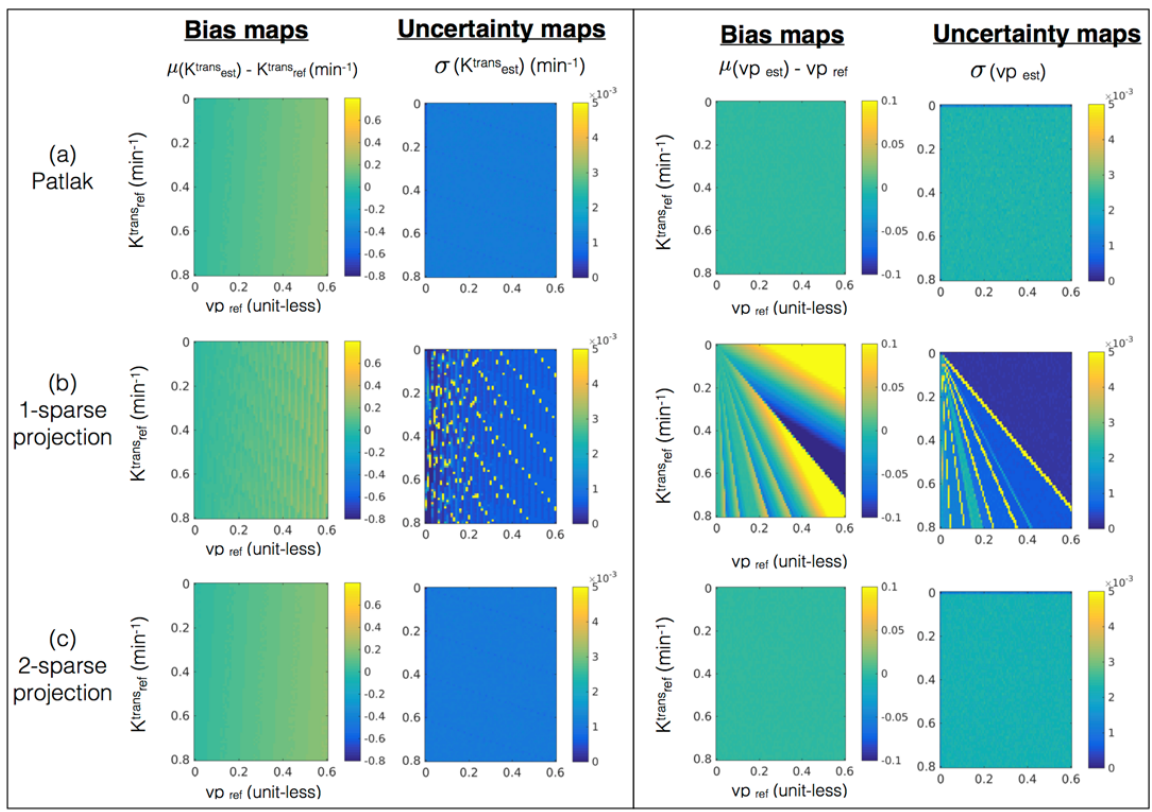

Figure 4: Error statistics (bias and uncertainty) in estimating kinetic parameters in presence of noise with the Patlak model. The first row in (a) shows the bias and uncertainty in estimating kinetic parameters from the noisy concentration vs time profiles and is considered as reference. Rows (b) and (c) show the bias and uncertainty in kinetic parameter estimation after q-sparse projection of the noisy profiles with different values of q, and is evaluated against the reference. It can be seen from (a) that q=1 demonstrates considerable bias. However, when q=2, the bias and uncertainty maps are equivalent to the reference, which motivated the choice of q=2 for the Patlak model.

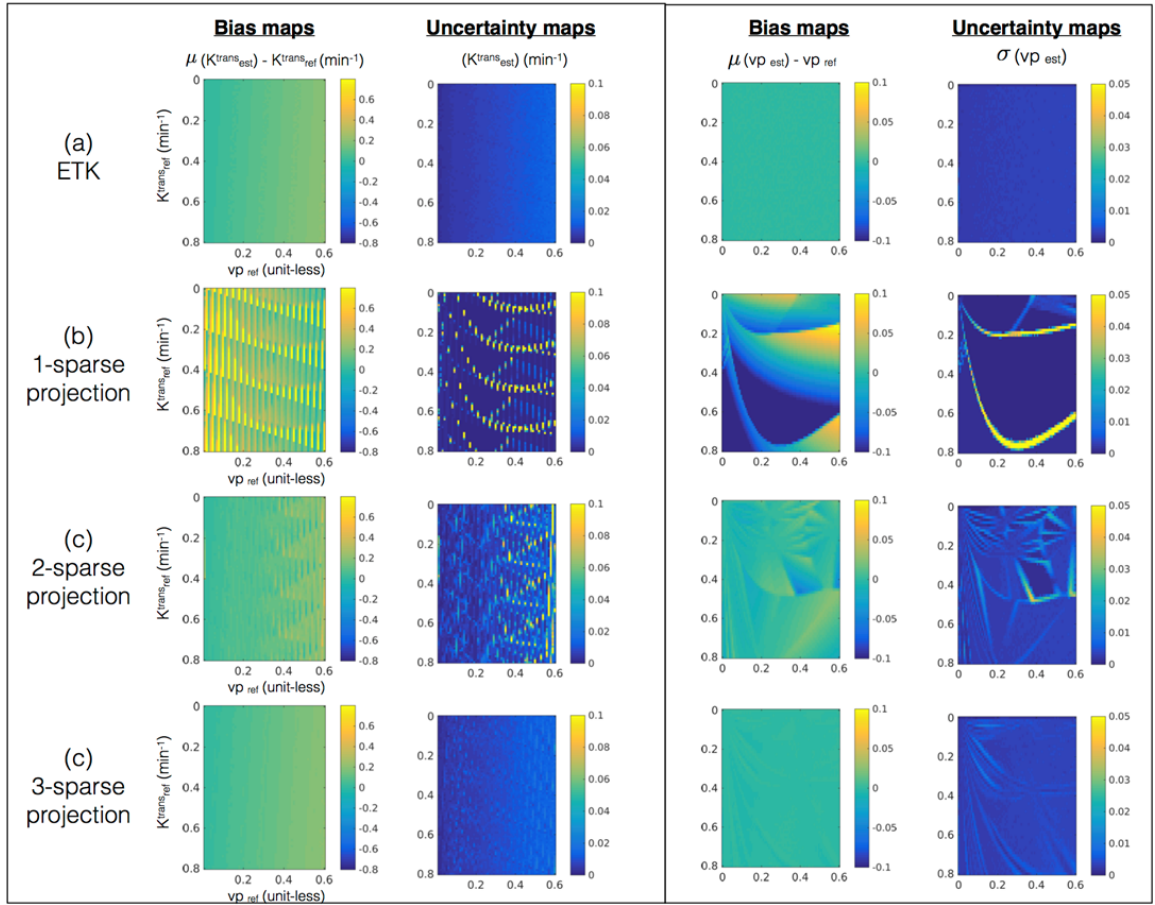

Figure 5: Error statistics (bias and uncertainty) in estimating kinetic parameters in presence of noise with the ETK model. The first row in (a) shows the bias and uncertainty in estimating kinetic parameters from the noisy concentration vs time profiles and is considered as reference. Rows (b-d) show the bias and uncertainty in kinetic parameter estimation after q-sparse projection of the noisy profiles with different values of q, and is evaluated against the reference. It can be seen from (a) and (b) that q=1, and q=2 demonstrates considerable bias and uncertainty. However, when q=3, the bias and uncertainty maps are similar to the reference over a broad range of the parameter space. This motivated our choice of q=3 for the ETK model.

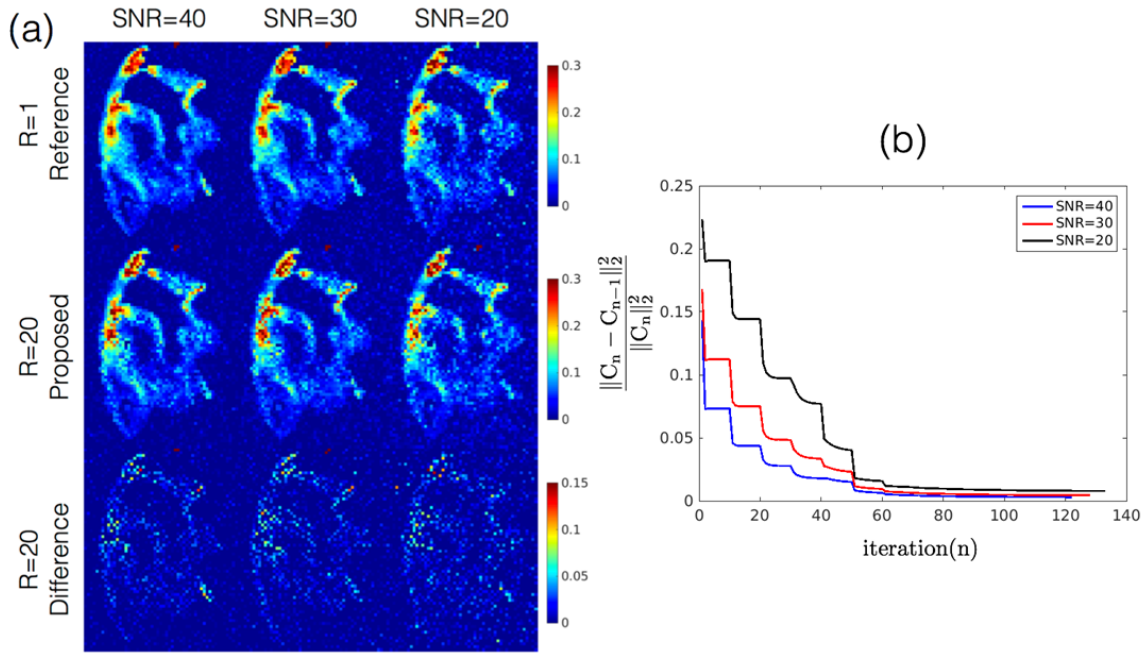

Figure 6: Evaluations using a digital reference object (DRO): (a) Comparison of $K^{trans}$ maps at SNR=40, 30, 20 obtained from the reference fully sampled (first row), and the proposed reconstruction at R=20 (second row). The difference maps are shown in the third row. (b) shows the per iteration change in the estimate of the concentration profiles over iteration number. From (b), the convergence across the different SNR settings depicts a similar pattern, but with different convergence speed. At convergence, the number of iterations for SNR=40, 30, 20 respectively were 122, 128, 134. As the SNR is decreased, the $K^{trans}$ estimates depict noise in both the reference and the proposed approach. The proposed approach at R=20 depicts good spatial fidelity of the $K^{trans}$ maps at all the noise levels ad also highlighted in the difference maps.

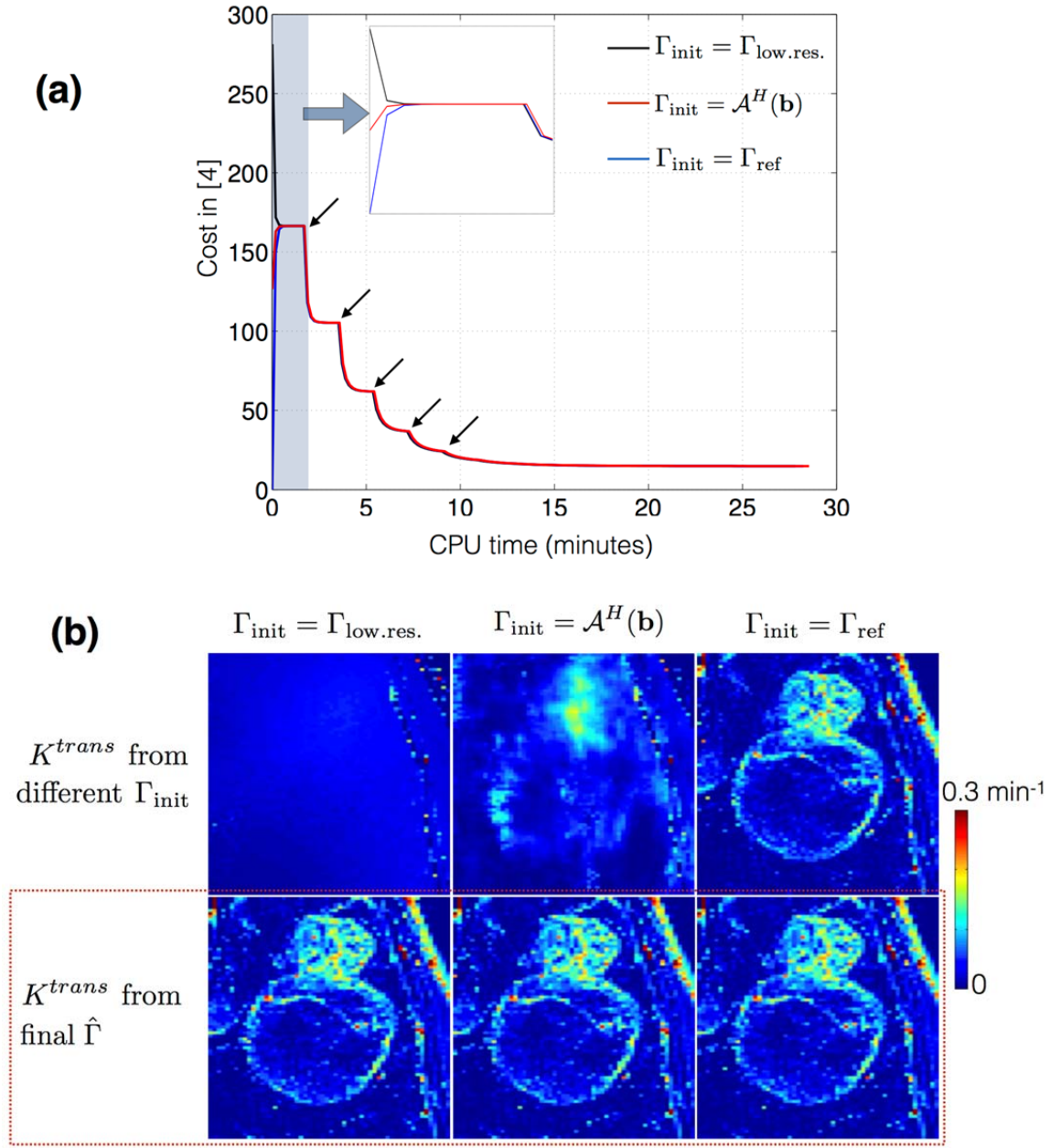

Figure 7: Convergence of cost function in [4]: (a) shows the evolution of cost with different initializations of the concentration time profiles ($\Gamma$). (b) demonstrates the $K^{trans}$ estimated from the initial guesses (top row), and from the final estimated concentration time profiles (bottom row). Due to the iterative multi-scale optimization, the algorithm ensures cycling through problems of increasing complexity. The black arrows in (a) indicate the instances at which the scale (spatial resolution) is incremented. It can be seen in (a) that the cost converges to the same minima irrespective of the initializations. The final estimated $K^{trans}$ in (b) from the different initializations are identical (see red dotted box).

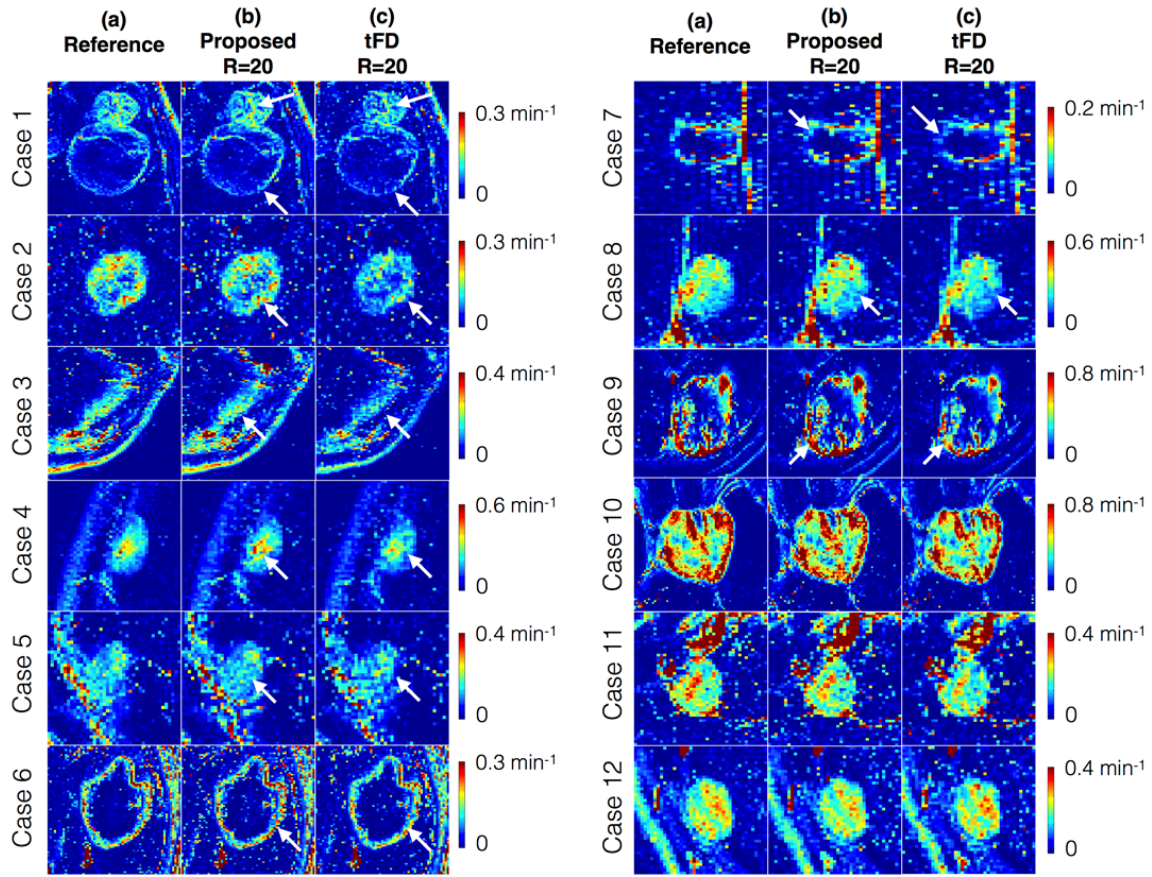

Figure 8: Evaluation of K$^{trans}$ maps derived from the proposed dictionary based and tFD reconstructions (R=20) against the reference K$^{trans}$ maps (R=1). The 12 cases are sorted based on decreasing difference between the proposed and tFD methods. The tFD reconstructions demonstrated under-estimation of K$^{trans}$ (visually evident in cases 1 to 9, see arrows). tFD also relied on tuning of a regularization parameter. In contrast, the proposed parameter-free model-based reconstruction provided better accuracy in K$^{trans}$ estimation, and also has improved fidelity in preserving spatial characteristics of the tumors (eg. thin boundaries of the tumor, see arrows in cases 1-5).

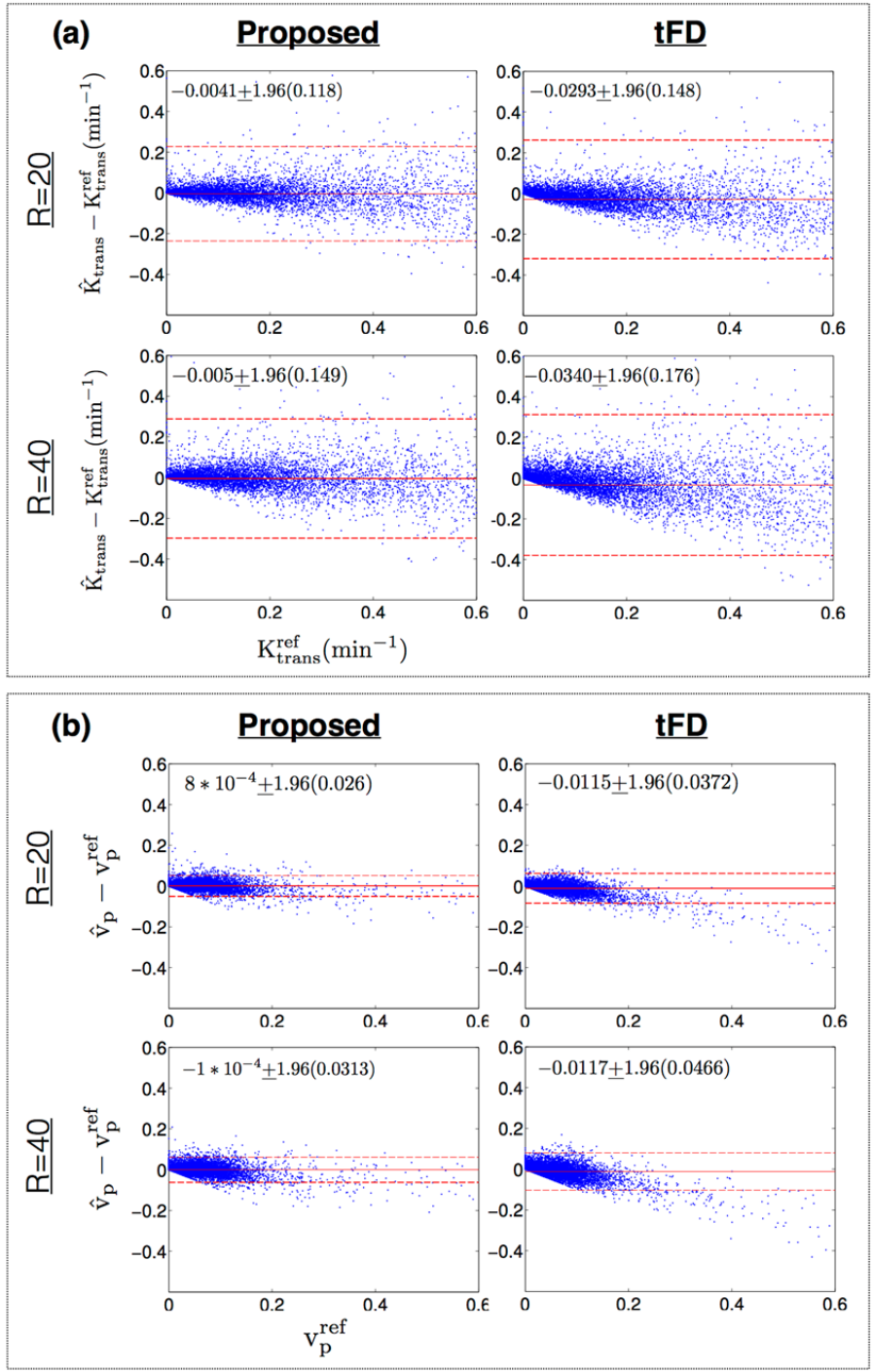

Figure 9: Bland-Altman plots of a) the difference between estimated K$^{trans}$ (at R=20, R=40) and reference K$^{trans}$; b) the difference between estimated $v_p$ (at R=20, R=40) and reference $v_p$; for the proposed (left column) and tFD (right column) reconstructions. Each dot corresponds to one pixel within the tumor ROIs of all the 12 cases. The mean and 1.96 times the standard deviation (μ+1.96σ) of the difference entities are quantitatively shown. These are also qualitatively marked by the solid red and dotted red lines. As seen from the plots, the proposed approach had lower bias ($\mu$) and uncertainty ($\sigma$) in estimating $K^{trans}$, and $v_p$ in comparison to tFD. tFD depicted a systematic bias in under-estimating $K^{trans}$, and $v_p$ in comparison to the proposed approach This can also be noted from the qualitative comparisons in Figure. 8.